\documentclass[final]{siamltex}

\usepackage{epsfig}
\usepackage{longtable}
\usepackage{graphicx,cite}
\usepackage{amsmath}

\def\bea{\begin{eqnarray}}
 \def\eea{\end{eqnarray}}

\begin{document}

\title{\bf DISCRETE MULTISCALE ANALYSIS: \\ A BIATOMIC LATTICE SYSTEM 
\footnote{This is a revised version of the article  in Journal of Nonlinear
  Mathematical Physics  Volume: 17, Issue: 3(2010) pp. 357-377     DOI: 10.1142/S1402925110000957}}


\author{ G.A. Cassatella Contra,\thanks{
Departamento de F\'isica Te\'orica II (M\'etodos Matem\'aticos de la
F\'isica), Universidad Complutense de Madrid,
Ciudad Universitaria, 28040 - Madrid, Spain}\and
D. Levi\thanks{
Dipartimento di Ingegneria Elettronica, 
Universit\`a degli
 Studi Roma Tre and Sezione INFN Roma Tre,
Via della Vasca Navale 84,
 00146 Roma, Italy}}

\maketitle


\begin{abstract}

   We discuss a discrete approach to the multiscale reductive
  perturbative method and apply it to a biatomic chain with a nonlinear
  interaction between the atoms. This system is important to describe the time
  evolution of localized solitonic excitations.
  
  We require that also the reduced equation be discrete. To do so coherently we need to discretize the time variable to be able to get asymptotic discrete waves  and carry out a discrete multiscale expansion around them. Our resulting
  nonlinear equation will be a kind of discrete Nonlinear Schr\"odinger
  equation. If we make its continuum limit, we obtain the standard  Nonlinear
  Schr\"odinger differential equation.
\end{abstract}

\pagestyle{myheadings}
\thispagestyle{plain}
\markboth{G.A. Cassatella Contra and D. Levi}{DISCRETE MULTISCALE ANALYSIS}

\section{INTRODUCTION}
\indent Nonlinear systems, and in particular nonlinear discrete systems, are gaining
  an increasing impact in modern science \cite{scott}.  

In 1955 Fermi, Pasta and Ulam (FPU) \cite{fermi} considered a
unidimensional chain of atoms with nonlinear nearest neighbouring interaction
to verify if nonlinearity could produce energy equipartition. Instead, they
found  recurrence, i.e. the motion of the chain for small energies
 was almost periodic \cite{50years}.  To explain this result Kruskal and
Zabusky found in 1965 \cite{kz} a connection between the FPU
system and the Korteweg--De Vries equation (KdV), an equation introduced in
fluid dynamics to describe one dimensional surface waves in the shallow
water context \cite{devries}.  By introducing the Inverse Scattering
Transform, they where able to solve the Cauchy problem for the KdV equation
\cite{skdv} and to prove the existence of soliton solutions.

In 1967 Toda \cite{fondamentale} considered a dynamical system with exponential interaction,
$U(r)=e^{-r}+r-1$, the "Toda potential", whose small amplitude approximation
gives the FPU system, and shares many of the integrability properties of the
KdV equation. So the FPU system turns out to be an approximation of a discrete soliton model.

Later more complicate atomic chains have been considered, as, for example, the
biatomic one \cite{bransden,henry,dash,dash1,Mokross,campa}.  These systems
have various applications in physics and biology as, for example, in the study
of ferroelectric perovskites, materials that, in certain crystallographic
directions have an almost unidimensional frame, and in organic molecular
chains.  A biatomic chain of neighboring atoms $A_1$ and $A_2$ is described by
the discrete independent variable $n$ and a continuous time $t$.  However, the
simplest nonlinear coupled lattice dynamical equations one can construct for
this system are not solvable. Only special exact solutions may be found.

Multiscale expansions \cite{devries,6,7,19,25,38,39} have proved to be
important tools to find approximate solutions for many physical problems by
reducing a given nonlinear partial differential equation to a simpler
equation, which is often integrable \cite{orszag}. Recently, few attempts to
carry over this approach to partial difference equations have been proposed
\cite{2,26,28,ckv,schoombie}. Almost all approaches considered contain some
approximation, either based on physical or on mathematical reasoning as
scaling transformations of the lattice provide a nonlocal result.  In the
following we prefer to stick to mathematical approximations as in this case it
will be more evident what to do to improve the final result \cite{hlps}.

 In ref \cite{campa} a biatomic chain obtained as a first nonlinear
 approximation of a complex Lenard--Jones interaction between atoms has been
 considered.  There the multiscale expansion of the continuous limit of the
 lattice model showed that the modulation of periodic solutions is governed by
 the Nonlinear Schr\"odinger differential Equation (NLSE). Here we consider
 the same model but we are interested in carrying out the multiscale expansion
 on the lattice, i.e. we are looking for a lattice equation which in the
 asymptotic regime approximate the biatomic nonlinear lattice. To do so we
 need to discretize time to be able to allow for discrete asymptotic waves. If
 we keep a continuous time variable an asymptotic wave travelling on the
 lattice by necessity will be described by a continuous variable. So by
 necessity we go over to a differential system.

Discretization of variables, besides representing an interesting problem in
mathematical physics for its computerizability, it is also useful in itself. Measurements, for example, are based on sampling of physical variables such as space
and time. It follows that physical models in which variables are defined on
the lattice are easier to be compared with the real world we see in our measurements.

In this work, we propose to continue the previous researches of
biatomic chains considering both $t$ and $n$ as discrete variables. In particular, we shall assume, as these authors, that the system
has a unharmonic cubic potential as in nature, potentials usually are
non--symmetric.  We shall thus 
apply a discrete multiscale reductive perturbative method to the model introduced by
Campa et. al.\cite{campa}
  consisting of a biatomic chain with a nonlinear nearest neighbour
  interaction.

In Section 2 we describe in detail the biatomic chain and write down the
dynamical equations. Then in Section 3 we introduce some notions of discrete
calculus and multiple scales defined on the lattice which we apply in Section
4 to the biatomic chain introduced in Section 2.  In Section 5 we analyze the
resulting nonlinear discrete equation obtained and carry out its continuum
limit. Finally, in Section 6 we draw some final conclusions.

\section{ THE MODEL}
\begin{figure}[ht]
\begin{center}
\includegraphics[scale=.75]{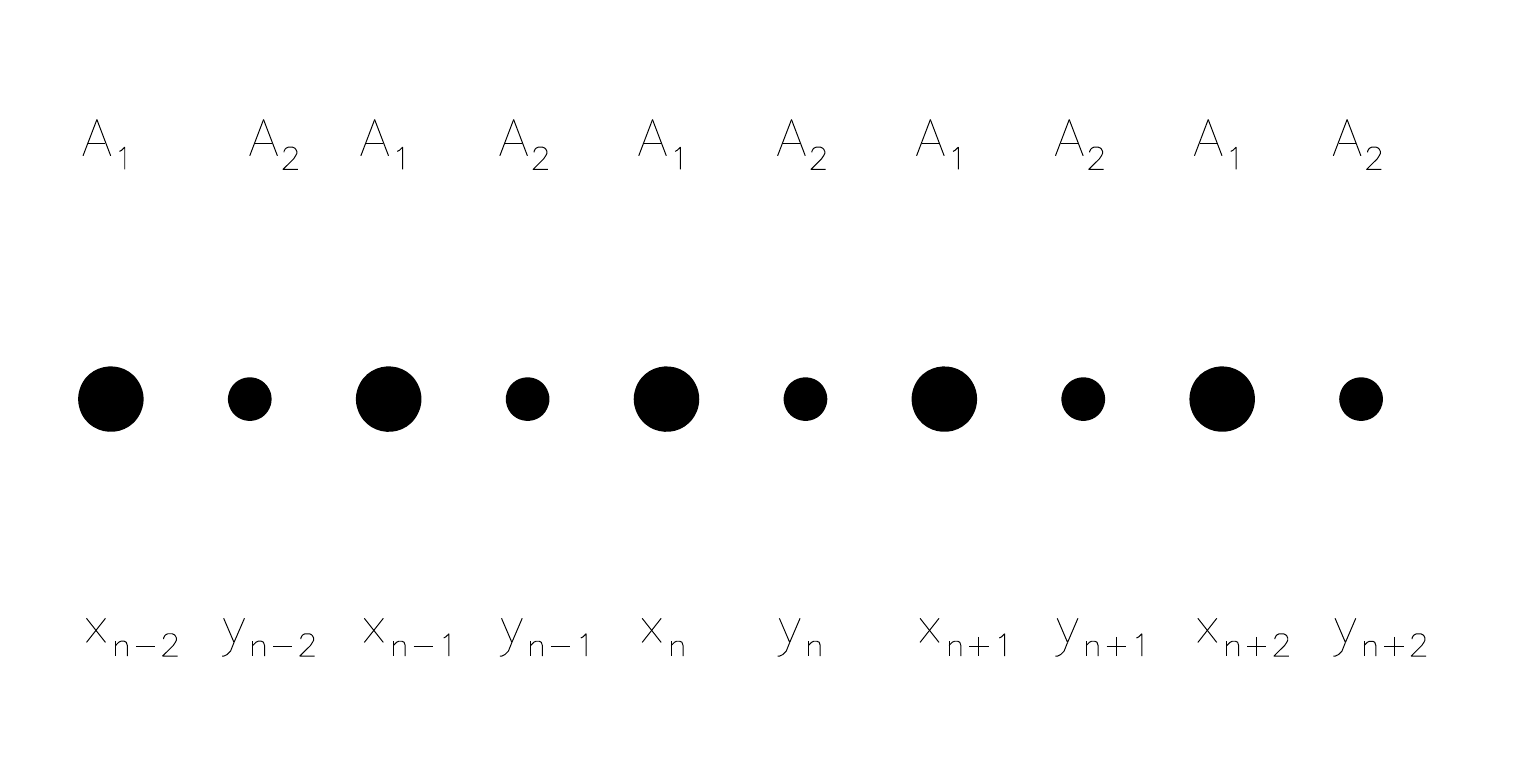}
\caption{\small Pattern of a biatomic molecular chain in one dimension. The chain is
  formed by a sequence of pairs of neighboring atoms $A_1$ and $A_2$. The
  displacements of the atoms of the molecule $n$ are indicated with $x_n$ and
  $y_n$.}
\label{fig:retfig}
\end{center}
\end{figure}

We want to describe here a chain suitable to represent, for example, an
$\alpha$-helix channel, see Scott
  (1999) \cite{scott}.
Our model consists of a biatomic chain formed by a sequence of pairs of
neighboring atoms $A_{1}$ and $A_{2}$, with masses $M_{1}$ and $M_{2}$,
respectively. Each pair, made  of an atom of mass $M_1$ and the following one of mass $M_2$, can be considered as a "molecule".  We denote by the index $n$ the $n^{th}$ molecule formed by the atom $A_{1}$ and $A_{2}$
(see Fig. \ref{fig:retfig}).
 Let us indicate with $x_{n}(t)$ and $y_{n}(t)$ the displacements  of the atoms $A_{1}$ and $A_{2}$ belonging to the same
 molecule $n$.  For each atom, we assume only nearest neighbourg
 interactions. Then, the total potential of the chain is given by
 \bea \label{a1} U=\sum_{n}\{U_{1}(y_{n}-x_{n})+U_{2}(x_{n+1}-y_{n})\}, \eea
 where $U_1$ is the intramolecular potential, between atoms belonging to the
 same molecule, and $U_2$ is the potential between different molecules. 

Given a natural \cite{bransden,ashcroft,toda1989} asymmetric potential with an
absolute minimum in the equilibrium position as, for example, a Lenard--Jones
potential, by taking the first terms of its Taylor expansion around the
equilibrium position we can write the potentials $U_1$ and $U_2$ as
\[U_{1}(r)=\frac{1}{2}k_{1}r^{2}
+\frac{\epsilon}{3}\beta_{1}r^{3},\qquad
U_{2}(r)=\frac{1}{2}k_{2}r^{2}
+\frac{\epsilon}{3}\beta_{2}r^{3},\] where $k_{1}$ and $k_{2}$ are the
harmonic constants, $\beta_{1}$ and $\beta_{2}$ are the cubic interaction
constants and $\epsilon$ is a small parameter which will play the role of the
perturbative parameter.  We assume that the
interaction between atoms of the same site is stronger than that of atoms of
different sites; thus $k_{1}${$>$}${k_{2}}$ and
$|\beta_{1}|${$>$}$|\beta_{2}|$.
So, the Hamiltonian of our molecular chain turns out to be
\[\begin{array}{c}
H=\sum_{n}^{}\{\frac{1}{2}[M_{1}\dot{x}_{n}^{2}+M_{2}\dot{y}_{n}^{2}]
+\frac{1}{2}[k_{1}(y_{n}-x_{n})^{2}+k_{2}(x_{n+1}-y_{n})^2]\\
+\frac{\epsilon}{3}[\beta_{1}(y_{n}-x_{n})^{3}+\beta_{2}(x_{n+1}-y_{n})^3]\},
\end{array}\]
 where $\dot{x}(t)${$\equiv$}$\frac{\text{d}x(t)}{\text{d}t}$ and the equations
of motion are
\bea \label{eq:1S}
M_{1}\ddot{x}_{n}&=&-\frac{{\partial}H}{{\partial}x_{n}}\\ \nonumber
&=&k_{1}(y_{n}-x_{n})-k_{2}(x_{n}
-y_{n-1})+\epsilon\beta_{1}(y_{n}-x_{n})^{2}
-\epsilon\beta_{2}(x_{n}-y_{n-1})^{2},
\\ \label{eq:2S}
M_{2}\ddot{y}_{n}&=&\frac{{\partial}H}{{\partial}y_{n}}\\ \nonumber
&=&-k_{1}(y_{n}-x_{n})+k_{2}(x_{n+1}
-y_{n})-\epsilon\beta_{1}(y_{n}-x_{n})^{2}
+\epsilon\beta_{2}(x_{n+1}-y_{n})^{2}.
\eea 

Eqs. (\ref{eq:1S}, \ref{eq:2S}) are a natural extension of the
FPU model \cite{fermi} to a biatomic system.

\section{ MULTIPLE SCALES ON A LATTICE} \label{s2}

Here we introduce the concepts necessary to extend the
multiscale reductive perturbative approach introduced by Poincar\'e \cite{orszag} for the
study of the asymptotic expansion of ordinary differential equations and
extended by Taniuti to the reduction of partial differential equations \cite{38,39} to the
case of difference equations \cite{schoombie,levi,hlps}.

\subsection{ Lattices and functions defined on  them}

Given a lattice, we will denote by $n$ the running index of the points
separated by a constant spacing $h$. Thus to the lattice {\bf index} $n$, we
can associate a {\bf continuous variable} $x$={$n$}{$h$} defining the position
of the points with respect to the origin, for convenience chosen to be with no
loss of generality $x_0=0$.

\begin{figure*} 
\begin{center}
    \includegraphics[scale=.8]{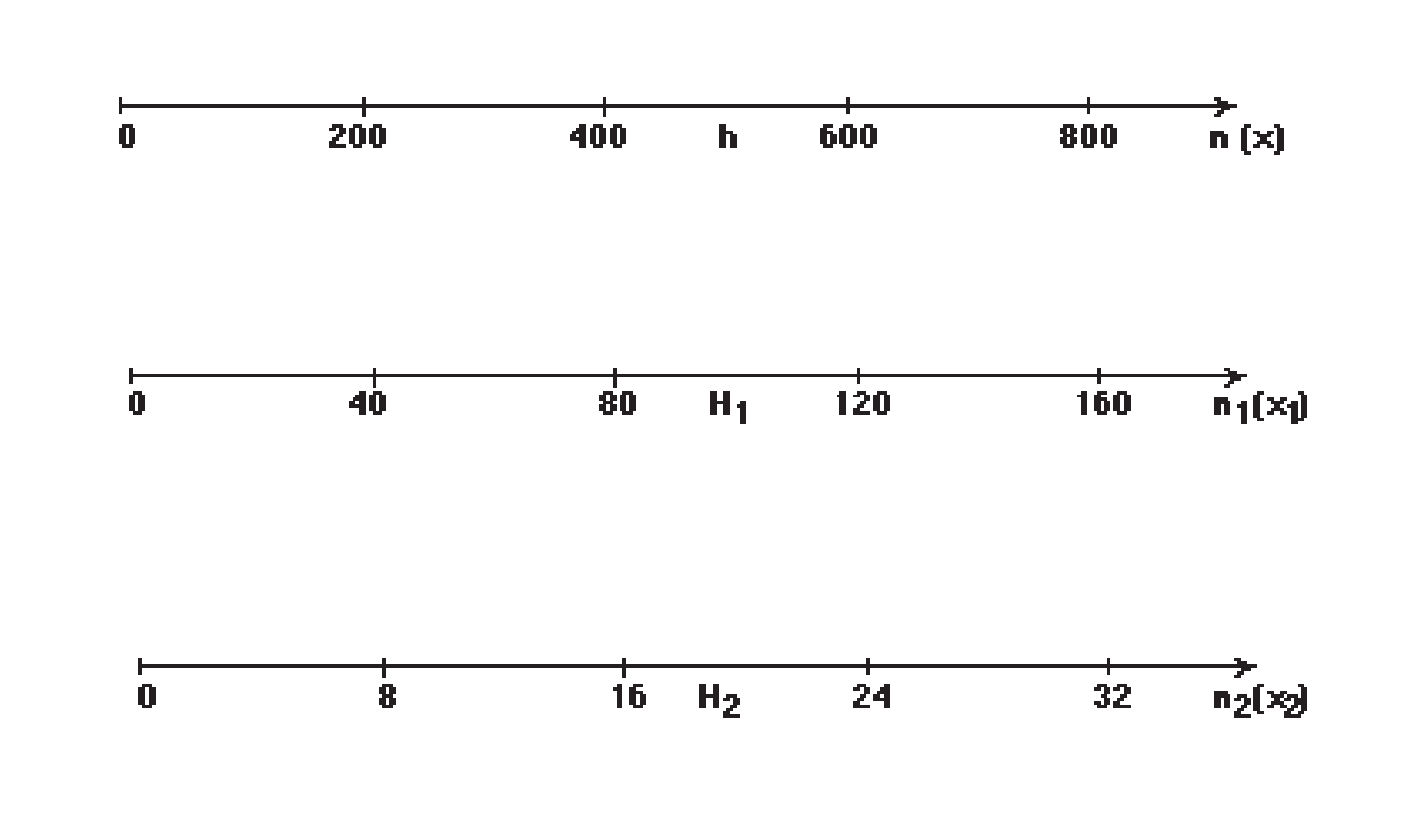}
    \caption{\small Rescaled lattices.} 
\label{fig:lat} 
\end{center}
 \end{figure*}

If we introduce a small parameter  $\epsilon$=$N^{-1}$, where $N$ is a large
integer positive number, we can define on the same lattice the slowly varying
discrete variables $n_{j}$ ($j$=1,2,3,...) of constant spacing $H_j$ and the continuous variables $x_j$ (see Fig. \ref{fig:lat}) where
\begin{equation}
  n=N^j n_{j}, \qquad x_j = \epsilon^j x.
  \label{eq:prodi}
\end{equation}

If $n_{j}$ varies by one, $n$ varies by $N^{j}$, a number much larger than
 unity. For this reason, $n_{j}$ is a ``slow variable'' and provide an
 asymptotic behavior of the system.  For each $j$ there is a slow lattice
 variable corresponding to the slow index $n_{j}$.  $n_{j}$ will be an integer
 only if $n$ is a common multiple of $N^{j}$.

 Let us consider $F_{n}$, a function of the discrete index $n$.  An equation
 on the lattice is a functional relation which involves the function $F$ at
 various lattice points, $\{F_{n+\ell}\}$. In the case of the model considered
 before (\ref{eq:1S}, \ref{eq:2S}), $\ell = \pm 1$.  We are interested to
 transform the system, defined on a lattice $n$, to the slowly varying
 lattices $n_j$, providing the scales of the asymptotic behavior of the
 original system. This is equivalent to say that we are interested in
 transforming the system defined on $x$ to the one with the slowly varying
 variables $x_j$.  We can consider the function $F_n$ written in terms of the
 slowly varying lattice variables $\{n_j\}$, with, for example, $j=1,2$, $ F_n \equiv f_{n_1,n_2}, $ and we can carry out the $\epsilon$
 expansion of the function $F_{n+\ell}$.

 Let us consider at the beginning the case of one slowly varying lattice
 $n_1$, i.e. $F_n\equiv f_{n_1}$. As the shift operator $T_n$ acting on $F_n$
 gives $ T_n F_n = F_{n+1}, $ we have
 $F_{n+\ell}=T_n^{\ell} F_n$. In order to extract the behaviour of the function $F_{n+1}=F(x+h)$ on the new scales, let us carry out the Taylor
 expansion of $F_{n+1}$ in powers of $h$. In such a case the shift operator can be expressed
 as an infinite order differential operator with respect to $x$, i.e.  \bea
\label{e2.3} T_n=\exp(h\partial_x)=\sum_{k=0}^{\infty}
\frac{(h \partial_x)^k}{k!}.  \eea Moreover, if we define a $\sl \Delta$
operator as $\Delta^{(+)}_n \equiv (T_n-1)/h$, we have \bea \label{e2.5a}
\partial_x=\frac{\log(1+h\Delta^{(+)}_n)}{h}, \eea and eq. (\ref{e2.3}) could
be written as \bea \label{e2.5ab} T_n=\sum_{k=0}^{\infty} \frac{(\log(1+h
  \Delta^{(+)}_n))^k}{k!}.  \eea Formulas (\ref{e2.5a}, \ref{e2.5ab}) are
written in terms of $\Delta_n^{(+)}$. However on the lattice we can define an
infinite number of different difference operators which in the continuum
limit, when $h$ goes to zero, go over to the first order derivative. Among
them it is important the asymmetric shift operator $\Delta_n^{(s)}\equiv
\frac{1}{2h}(T_n-T_n^{-1})$. In this case we have \bea
\label{eq:romauno}
\partial_x =\frac{\mathrm{arcsinh}(h \Delta_n^{(s)})}{h}, \quad \rightarrow
\quad T_n=\sum_{k=0}^{\infty} \frac{(\mathrm{arcsinh}(h
  \Delta_n^{(s)}))^k}{k!}.  \eea Introducing the slowly varying variable $x_1$
and the corresponding lattice $n_1$ in eq. (\ref{e2.3}), as $\partial_x =
\epsilon \partial_{x_1}$, we have \bea \label{e2.4} T_n^{\ell} = e^{\ell
  h\partial_x} = e^{\ell \epsilon h\partial_{x_1}} = T_{n_1}^{\ell \epsilon}
=\sum_{k=0}^{\infty} \frac{(\ell h \epsilon \partial_{x_1})^k}{k!}. \eea If we
introduce more lattice variables, for example $\{n_j\}$, with $j=1,2$, then
$T_n$ becomes \bea \label{e2.4a} T_n^{\ell} = T_{n_1}^{\ell \epsilon}
T_{n_2}^{\ell \epsilon^2}=\sum_{k=0}^{\infty} \frac{(\ell \epsilon h
  \partial_{x_1})^k}{k!}  \sum_{j=0}^{\infty} \frac{(\ell \epsilon^2 h
  \partial_{x_2})^j}{j!}. \eea Once we expand the operator $\partial_{x_j}$ in
terms of shift operators we get an expression for $F(n \pm \ell)$ in terms of
variations of $f(n_1,n_2)$ with coefficients depending on $\epsilon$ and
$\ell$.

 As delta operators are linear combinations of shift operators, from
 eq. (\ref{e2.5c}) it can be proved \cite{jordan,levi}  that for
 $\Delta=\Delta^{(+)}$ we have the following formula
\begin{equation}
  ({\Delta}^{(+)}_{n_{1}})^{k}f_{n_{1}}=\sum_{i=k}^{\infty}\frac{k!}
{i!}P(i,k)({\Delta}_n^{(+)})^{i}F_{n},
\label{eq:delta}
\end{equation}
where $({\Delta}_{n_{1}}^{(+)})^{k}f_{n_{1}}$ is the $k$th-difference of
$f_{n_{1}}$ respect to $n_{1}$, and the coefficients $P(i,k)$ are given by
$P(i,j)=\sum_{\alpha=j}^{i}w^{\alpha}S_{i}^{\alpha}G_{\alpha}^{j},$ where
$\omega$ is the ratio of the increment in the lattice of variable $n$ with
respect to that of variable $n_{1}$. In this case, taking into account
equation (\ref{eq:prodi}), $\omega$=$N$. The coefficients $S_{i}^{\alpha}$ and
$G_{\alpha}^{j}$ are the Stirling coefficients of the first kind and second
kind, respectively. The result (\ref{eq:delta}) can be inverted, providing:
\begin{equation} ({\Delta}_n^{(+)})^{k}F_{n}=
  \sum_{i=k}^{\infty}\frac{k!}{i!}Q(i,k)({\Delta}^{(+)}_{n_{1}})^{i}f_{n_{1}},
\label{eq:nabla}
\end{equation}
where $Q(i,j)$ is the same as $P(i,j)$, but with $w =N^{-1}=\epsilon$.

A general way to get these formulas is provided by the {\sl finite operator
  calculus} \cite{rota,roman,revu}.  The finite operator calculus prescribes
the following formula \cite{LT} \bea \label{e2.5c} T_n^j= \sum_{k=0}^{\infty}
\frac{(\epsilon)^k p_k(j)}{k!} (\Delta_{n_1})^k, \eea where the functions
$p_k(j)$ are the unique basic sequence associated to the operator
$\Delta_{n_1}$, i.e. such that they satisfy the following conditions \bea
\label{e2.5d} &&p_0(n_1)=1, \qquad p_k(0) = 0 \quad \mbox{for all} \quad k>0,
\\ \nonumber &&\Delta_{n_1} p_k(n_1) = k p_{k-1}(n_1).  \eea The basic
sequences can be directly obtained by the transfer formulae: \bea
  \label{e2.5e} p_k(n_1) = n_1 \bigl( \frac{\Delta_{n_1}}{h\partial_{x_1}}
  \bigr)^{-k} n{_1}^{k-1}.  \eea When $\Delta_{n_1} = \Delta_{n_1}^{(+)}$ or
  $\Delta_{n_1} = \Delta_{n_1}^{(s)}$, the basic sequences are: \bea
  \label{e2.5f} &&p_k^{(+)}(n_1) = h^{k}n_1 \bigl(
  \frac{e^{h\partial_{x_1}}-1}{h\partial_{x_1}}\bigr)^{-k} n{_1}^{k-1} =
  (x_1)_k \equiv x_1 (x_1-h) \cdots (x_1-kh+h), \\\nonumber&&p_k^{(s)}(n_1)
  =h^{k}n_1\bigl(\frac{e^{h\partial_{x_1}}
-e^{-h\partial_{x_1}}}{2h\partial_{x_1}}\bigr)^{-k}
  {n_1}^{k-1} = (2h)^k G_k(x_1;-h,2h), \eea where $G_k(y;a,b)$ are the Gould
  polynomials \cite{roman} given by 
  \bea \label{eq:e2.5g}
  G_k(y;a,b)&{\equiv}&\frac{y}{y-ka}
 \bigl(\frac{y-ka}{b}\bigr)_k\\ 
  &=&\frac{y}{(y-ka)(b)^k}(y-ka)(y-ka-b)\cdots(y-ka-(k-1)b).
\nonumber
\eea
Let us also mention that for each $\Delta_{n_1}$ operator we
  can write from eq. (\ref{e2.5c}) \bea
  \label{e2.5h} (\partial_{x_1})^j = \frac{1}{h^{j}}\sum_{k=0}^{\infty} \frac{1}{k!}
  \bigl[ \frac{\text{d}^j}{\text{d}y^j}p_k(y) \bigr] {\bigr |}_{y=0}
  (\Delta_{n_1})^k, \eea i.e. we can express the partial derivative as an
  infinite sum of differences whose coefficients depends from the type of
  difference we are expanding into. In terms of $\Delta^{(+)}$, from
  eqs. (\ref{e2.5c}, \ref{e2.5f}), eq. (\ref{e2.4}) reads \bea \label{e2.51i}
  T_n^{\ell} F_n= \sum_{k=0}^{\infty} \frac{(h\epsilon)^k(\ell)_k}{k!}
  ({\Delta}^{(+)}_{n_{1}})^{k} f_{n_1},  \eea while, in the symmetric
  difference case, it reads
\begin{equation}
T_n^{\ell}F_n=\sum_{k=0}^{\infty}\frac{(2h\epsilon)^{k}}{k!}
G_k(l;-1,2)({\Delta}^{(s)}_{n_{1}})^{k}f_{n_1}.  
\label{eq:e2.5i} 
\end{equation}
From eqs. (\ref{e2.51i},\ref{eq:e2.5i}) we get that any finite shift in the original equation
will give rise to an expression in the slowly varying variables which involves
an infinity of lattice points or, equivalently, contains differences at all
orders of the function $f_{n_1}$. So to get a reduced equation on a finite
number of points we need to cut the series by requiring that the function
$f_{n_1}$ be of finite order of variation. Let us introduce the following
definition:

$\mathbf{Definition:}$
{\it The function $f_{n}$ is a slow varying function of order $p$ if
\begin{equation}
\Delta^{p+1}f_{n}=0.
\label{eq:dim}
\end{equation}}
Than we can  prove the following Theorem:

$\bf{Theorem}$ {\it The function $F_{n}$ is a slow varying function of order $p$
iff $f_{n_1}$ is a slowly varying function of order $p$ in its own variable,
i.e. if $\Delta_{n_1}^{p+1} f_{n_1}=0$.}
\begin{proof}
The proof of this theorem will be given in the case of $\Delta=\Delta^{+}$, but it is easy to see that is valid for
any delta operator. It is divided in two parts:

\begin{description}
\item{\bf a)}
Be $f_{n_{1}}$ a slowly varying function of order $p$. From formula
(\ref{eq:nabla}) it follows that
\begin{equation}
{\Delta}_n^{p+1}F_{n}=\sum_{i=p+1}^{\infty}\frac{(p+1)!}{i!}Q(i,p+1)
{\Delta}_{n_{1}}^{i}f_{n_{1}}=0,
\end{equation}
i.e. $F_n$ is also a slow function of order $p$.
\item{\bf b)}
  Be $F_{n}$ a slowly varying function of order $p$. From formula
  (\ref{eq:delta}) it follows that
\begin{equation}
{\Delta}_{n_{1}}^{p+1}f_{n_{1}}=
\sum_{i=p+1}^{\infty}\frac{(p+1)!}{i!}P(i,p+1){\Delta}_n^{i}F_{n}=0,
\end{equation}
i.e. $f_{n_1}$ is also a slow function of order $p$.  \qquad ~
\end{description}\end{proof}

The expansion (\ref{eq:e2.5i}) can be performed in two steps: at first we write
the shift operator in the $n$ variable in terms of the derivatives with
respect to $x_1$ by formula (\ref{e2.4}) and then we expand the derivatives
with respect to $x_1$ in term of delta operators by formula (\ref{e2.5h}). In
doing so we will have formulas in derivatives which are valid for any delta
operator. Moreover the first expansion has $\epsilon$ dependent coefficients
while the second will provide a finite number of terms only if we use the slow
varying condition for the functions $f_{n_1,n_2}$.

Let us now explicitate the first terms of eq. (\ref{eq:e2.5i}) for future use,
at first in term of the derivatives and then in delta operators assuming that
the function $f_{n_1,n_2}$ is a slow function at most of order 2.
At first we shall consider the case in which we have only one slow
lattice, just the variable $n_1$ is present and then we extend the result to
the case of two slow lattices, $n_1$ and $n_2$ and to partial lattices $n$ and $m$.

\subsubsection{$F_{n}$=$f_{n_{1}}$=$f(x_1)$}
From Eq. (\ref{e2.4}) we get \bea \label{isolafamosi0} F_{n\pm{1}}=f(x_{1})
\pm h \epsilon \partial_{x_1} f(x_{1}) +
\frac{h^{2}\epsilon^2}{2!}\partial^2_{x_1} f(x_{1}) + \mathcal O (\epsilon^3).
\eea As from eq.(\ref{e2.5h}) for $p$=2, $\partial_{x_1}$=$\Delta_{n_{1}}$ and
${\partial^2}_{x_1}$=$(\Delta_{n_{1}})^2$, then equation (\ref{isolafamosi0})
reads
\begin{equation}
F_{n\pm{1}}=f_{n_{1}}\pm\frac{1}{2N}(f_{n_{1}+1}
-f_{n_{1}-1})+\frac{1}{2N^2}(f_{n_{1}+1}-2f_{n_{1}}+f_{n_{1}-1})+O(N^{-3}).
\label{eq:isolafamosi}
\end{equation}

\subsubsection{$F_{n}$=$f_{n_{1},n_{2}}$=$f(x_1,x_2)$}
$p$=2 is the lowest nontrivial value of $p$ for
which we can consider $F_{n}$ as a function of the two scales, $n_{1}$ and
$n_{2}$.
Taking $l$=1, from eq.  (\ref{e2.4a}) we have
\bea \label{eq:consiglioministri}
  F_{n\pm1}&=&f(x_{1},x_{2})\pm
{h\epsilon}\frac{{\partial}f(x_{1},x_{2})}{\partial{x_1}}
  +\frac{h^2\epsilon^2}{2}\frac{{\partial^2}f(x_{1},x_{2})}
{{\partial}x_{1}^2}\pm{h\epsilon^2}\frac{{\partial}f(x_{1},x_{2})}{\partial{x_2}}\\ \nonumber
   &+&{h^2\epsilon^3}\frac{\partial}{\partial{x_1}}
\frac{{\partial}f(x_{1},x_{2})}{\partial{x_2}}+O(\epsilon^{4}).
\eea
If $F_{n}$ is a slowly varying function of order two in $n_{1}$, it might be of
order one in $n_{2}$. In this case, eq. (\ref{eq:consiglioministri})
becomes
\bea
  F_{n\pm1}=f(x_{1},x_{2})\pm
{h\epsilon}\frac{{\partial}f(x_{1},x_{2})}{\partial{x_1}}
  +\frac{h^2\epsilon^2}{2}\frac{{\partial^2}f(x_{1},x_{2})}
{{\partial}x_{1}^2}\pm{h\epsilon^2}
\frac{{\partial}f(x_{1},x_{2})}{\partial{x_2}}+O(\epsilon^{3}).
\label{eq:portaaporta}
\eea
Moreover, from eq. (\ref{e2.5h}) it follows  that $\partial_{x_2}$=$\Delta_{n_{2}}$,
$\partial_{x_1}^2$=$(\Delta_{n_{1}})^2$ and $\partial_{x_1}\partial_{x_2}$
=$\Delta_{n_{1}}\Delta_{n_{2}}$. Then eqs. (\ref{eq:consiglioministri},\ref{eq:portaaporta}), written in terms of differences instead of
derivatives, are given by
\bea \label{eq:spring}
  F_{n\pm1}&=&f_{n_{1},n_{2}}\pm\frac{1}{2N}(f_{n_{1}+1,n_{2}}
-f_{n_{1}-1,n_{2}}) \\ \nonumber
  &+&\frac{1}{2N^{2}}(f_{n_{1}+1,n_{2}}-2f_{n_{1},n_{2}}+f_{n_{1}-1,n_{2}})
  \pm\frac{1}{2N^{2}}(f_{n_{1},n_{2}+1}-f_{n_{1},n_{2}-1})\\ \nonumber
&+&\frac{1}{4N^3}(f_{n_{1}+1,n_{2}+1}-f_{n_{1}-1,n_{2}+1}-f_{n_{1}+1,n_{2}-1}
+f_{n_{1}-1,n_{2}-1})+O(N^{-4})
\eea
and
\bea \label{eq:summer}
  F_{n\pm1}&=&f_{n_{1},n_{2}}\pm\frac{1}{2N}(f_{n_{1}+1,n_{2}}-f_{n_{1}-1,n_{2}})\\ \nonumber
  &+&\frac{1}{2N^{2}}(f_{n_{1}+1,n_{2}}-2f_{n_{1},n_{2}}+f_{n_{1}-1,n_{2}})
  \pm\frac{1}{2N^{2}}(f_{n_{1},n_{2}+1}-f_{n_{1},n_{2}-1})+O(N^{-3}),
\eea
respectively.

\subsubsection{$F_{n,m}$=$f_{n_{1},m_{1},m_{2}}$=$f(x_1,t_1,t_2)$}
 In this case we
have
\bea \label{eq:corrupt}
  F_{n,m\pm1}&=&f(x_{1},t_{1},t_{2})\pm
{\tau}\epsilon\frac{{\partial}f(x_{1},t_{1},t_{2})}{\partial{t_1}} \\ \nonumber
  &+&\frac{\tau^2\epsilon^2}{2}\frac{{\partial^2}f(x_{1},t_{1},t_{2})}
{{\partial}t_{1}^2}
\pm{\tau\epsilon^2}\frac{{\partial}f(x_{1},t_{1},t_{2})}{\partial{t_2}}
+O(\epsilon^{3}),
\eea
and
\begin{equation}
F_{n\pm{1},m}=f(x_{1},t_{1},t_{2})\pm{h\epsilon}
\frac{{\partial}f(x_{1},t_{1},t_{2})}{\partial{x_1}}
+\frac{h^2\epsilon^2}{2}\frac{{\partial^2}f(x_{1},t_{1},t_{2})}
{{\partial}x_{1}^2}+O(\epsilon^{3}).
\label{eq:country}
\end{equation}
In terms of differences, the last two equations are given by
\bea  \label{eq:autumn}
F_{n,m\pm1}&=&f_{n_{1},m_{1},m_{2}}\pm\frac{1}{2N}(f_{n_{1},m_{1}+1,m_{2}}-
f_{n_{1},m_{1}-1,m_{2}})\\ \nonumber &+&\frac{1}{2N^{2}}(f_{n_{1},m_{1}+1,m_{2}}-2f_{n_{1},m_{1},m_{2}}
+f_{n_{1},m_{1}-1,m_{2}})\\ \nonumber &\pm&\frac{1}{2N^{2}}(f_{n_{1},m_{1},m_{2}+1}
-f_{n_{1},m_{1},m_{2}-1})+O(N^{-3})
\eea
and
\bea \label{eq:winter}
F_{n\pm{1},m}&=&f_{n_{1},m_{1},m_{2}}\pm\frac{1}{2N}(f_{n_{1}+1,m_{1},m_{2}}
-f_{n_{1}-1,m_{1},m_{2}})\\ \nonumber &+&\frac{1}{2N^2}(f_{n_{1}+1,m_{1},m_{2}}
-2f_{n_{1},m_{1},m_{2}}+f_{n_{1}-1,m_{1},m_{2}})+O(N^{-3}).
\eea
For future use we can further rescale the lattice with some extra parameter by
defining $n_{1}=\frac{L_{1}n}{N}$, $m_{1}=\frac{L_{2}m}{N}$ and
$m_{2}=\frac{m}{N^{2}}$, where the order 1 parameters $L_{1}$ and $L_{2}$ are
divisors of $N$ and $N^2$ respectively if we require that $n_{1}$ and
$n_{2}$ be integer numbers. In this case, the equations (\ref{eq:corrupt}) and
(\ref{eq:country}) become
\bea \label{eq:uno}
  F_{n,m\pm1}&=&f(x_{1},t_{1},t_{2})\pm
{{\tau}L_2\epsilon}\frac{{\partial}f(x_{1},t_{1},t_{2})}{\partial{t_1}}
  +\frac{\tau^{2}L_{2}^2\epsilon^2}{2}\frac{{\partial^2}f(x_{1},t_{1},t_{2})}
{{\partial}t_{1}^2} \\ \nonumber &\pm&{\tau\epsilon^2}
\frac{{\partial}f(x_{1},t_{1},t_{2})}{\partial{t_2}}+O(\epsilon^{3}),
\eea
and
\bea
F_{n\pm{1},m}=f(x_{1},t_{1},t_{2})\pm{{h}L_1\epsilon}
\frac{{\partial}f(x_{1},t_{1},t_{2})}{\partial{x_1}}
+\frac{h^{2}L_{1}^2\epsilon^2}{2}\frac{{\partial^2}f(x_{1},t_{1},t_{2})}
{{\partial}x_{1}^2}+O(\epsilon^{3}).
\label{eq:dos}
\eea
Moreover, from equation (\ref{eq:uno}) we have
\begin{equation}
F_{n,m+1}-2F_{n,m}+F_{n,m-1}={\tau^{2}L_{2}^{2}\epsilon^2}\frac{
{\partial^2}f(x_{1},t_{1},t_{2})}{{\partial}t_{1}^2}+O(\epsilon^{3}).
\label{eq:tres}
\end{equation}
In terms of symmetric difference operators  these
equations can be written as
\bea \label{eq:republica}
F_{n,m\pm1}&=&f_{n_{1},m_{1},m_{2}}\pm\frac{L_{2}}{2N}(f_{n_{1},m_{1}+1,m_{2}}-f_{n_{1},m_{1}
-1,m_{2}})\\ \nonumber &+&\frac{L_{2}^{2}}{2N^{2}}(f_{n_{1},m_{1}+1,m_{2}}-2f_{n_{1},m_{1},m_{2}}
+f_{n_{1},m_{1}-1,m_{2}})\\ \nonumber &\pm&\frac{1}{2N^{2}}(f_{n_{1},m_{1},m_{2}+1}
-f_{n_{1},m_{1},m_{2}-1})+O(N^{-3}),
\eea
\bea \label{eq:bananera}
F_{n\pm{1},m}&=&f_{n_{1},m_{1},m_{2}}\pm\frac{L_{1}}{2N}(f_{n_{1}+1,m_{1},m_{2}}
-f_{n_{1}-1,m_{1},m_{2}})\\ \nonumber &+&\frac{L_{1}^2}{2N^2}(f_{n_{1}+1,m_{1},m_{2}}
-2f_{n_{1},m_{1},m_{2}}+f_{n_{1}-1,m_{1},m_{2}})+O(N^{-3})
\eea
and
\bea
F_{n,m+1}-2F_{n,m}+F_{n,m-1}&=&\frac{L_{2}^{2}}{N^2}(f_{n_{1},m_{1}+1,m_{2}}
-2f_{n_{1},m_{1},m_{2}}+f_{n_{1},m_{1}-1,m_{2}})\label{eq:raccomandatiditalia} \\ \nonumber &+&O(N^{-3}).
\eea
The last three equations will be used in the following Section to apply the
multiscale method to the biatomic lattice model we introduced in Section 2.

\section{MULTISCALE REDUCTION OF THE DISCRETE BIATOMIC SYSTEM}  \label{s3}

\subsection{Equations of motion}

In the equations of motion of the biatomic chain (see
eqs. (\ref{eq:1S},\ref{eq:2S})), the nonlinear terms (proportional
to $\beta_{1}$ and $\beta_{2}$) are of order $\epsilon$ respect to the
remaining terms, and thus we can use perturbative methods to look for
approximate solutions of $x_{n}(t)$ and $y_{n}(t)$. This has been done in 1993
by Campa et al. \cite{campa} using the multiscale perturbative method with
just the lowest order differential terms. In this way, performing at the same
time a multiscale expansion and a continuum limit they were able to reduce the
system to the NLSE  (\ref{eq:berluscon}).

Here we discretize time and look for completely discrete equations,
i.e. passing from the differential terms in the expansion (see
eqs. (\ref{isolafamosi0}, \ref{eq:consiglioministri}, \ref{eq:portaaporta},
\ref{eq:corrupt}, \ref{eq:country}, \ref{eq:uno}--\ref{eq:tres})) to
difference terms corresponding to the lowest order of slow varyness $p$,
i.e. to eqs.  (\ref{eq:isolafamosi}, \ref{eq:spring}, \ref{eq:summer}, \ref{eq:autumn}, \ref{eq:winter}, \ref{eq:republica}--\ref{eq:raccomandatiditalia}). To discretize time we replace
the time $t$ with a discrete variable $m$, so that $t${$\equiv$}$m\tau$, where
$\tau$ is the temporal scale. Thus, when $\tau$ reduces to an infinitesimal
quantity and $m$ approaches infinity in such a way that $t$ remains finite we
recover the continuous case.
We consider the simplest approximation of the second derivative by differences
 using a central difference so as to get a real dispersive relation.  The
discretized equations of motion are given by
\bea \label{eq:1} m_{1}(x_{n,m+1}-2x_{n,m}+x_{n,m-1}) &=& k_{1}(y_{n,m}-x_{n,m})-k_{2}(x_{n,m}
  -y_{n-1,m})\\ \nonumber  &+&\epsilon [\beta_{1}(y_{n,m}-x_{n,m})^{2}-
  \beta_{2}(x_{n,m}-y_{n-1,m})^{2}],\\
\label{eq:2}
 m_{2}(y_{n,m+1}-2y_{n,m}+y_{n,m-1}) &=& - k_{1}(y_{n,m}-x_{n,m})+k_{2}(x_{n+1,m}
-y_{n,m}) \\ \nonumber  
 &-&\epsilon [\beta_{1}(y_{n,m}-x_{n,m})^{2}-\beta_{2}(x_{n+1,m}-y_{n,m})^{2}], \eea where
$x_{n,m}${$\equiv$}$x_{n}(m\tau)$, $y_{n,m}${$\equiv$}$y_{n}(m\tau)$ and
$m_{1,2}${$\equiv$}$\frac{M_{1,2}}{\tau^{2}}$. We are looking for $x_{n,m}$
and $y_{n,m}$ as bounded solutions written as a modulation of the harmonic
wave solutions of the linearized equations which one obtains when setting
$\epsilon =0$. The harmonic waves are given by
\begin{equation}
  E_{n,m}=e^{i[kn-{\omega}(k)m]},
\label{eq:onda}
\end{equation}
with $\omega(k)$ real for any real value of $k$. The physical reason for
choosing harmonic waves is that the atoms of the chain make only small
oscillations around their equilibrium position.  When we introduce this ansatz
into equations (\ref{eq:1}) and (\ref{eq:2}), we realize at once that 
the solution of the nonlinear equations of motion can be represented as a
modulated linear combination of harmonic functions.

A solution of the linear part of  eqs. (\ref{eq:1}) and (\ref{eq:2})
($\beta_{1}$=$\beta_{2}$=0), written in terms of the harmonic waves
(\ref{eq:onda}), is given by
\bea \nonumber
 x_{n,m}=A \, E_{n,m}, \qquad
y_{n,m}=B\, E_{n,m}, \eea
where
\bea
  \frac{B}{A}=r{\equiv}\frac{k_{1}+k_{2}+2m_{1}(\cos{\omega(k)}-1)}{k_{1}
+k_{2}e^{-ik}}=\frac{k_{1}+k_{2}e^{ik}}{k_{1}+k_{2}
+2m_{2}(\cos{\omega(k)}-1)},
\label{eq:reldisp}
\eea with the dispersion relation 
\bea \label{eq:dispersion} \omega(k)={\scriptstyle \arccos\{1-\frac{1}{4m_{1}m_{2}}\bigl [(k_{1}+k_{2})(m_{1}+m_{2})
\pm\sqrt{(k_{1}+k_{2})^{2}(m_{1}+m_{2})^{2}
  -16k_{1}k_{2}m_{1}m_{2}{sin}^{2}\frac{k}{2}}\bigr ]\}}.  \eea 
    It
can be proved that the term inside the square root of the dispersion relation
is always positive, so that the argument of ``arccos'' is always real.

\begin{figure*}
\begin{center}
\includegraphics[width=8cm]{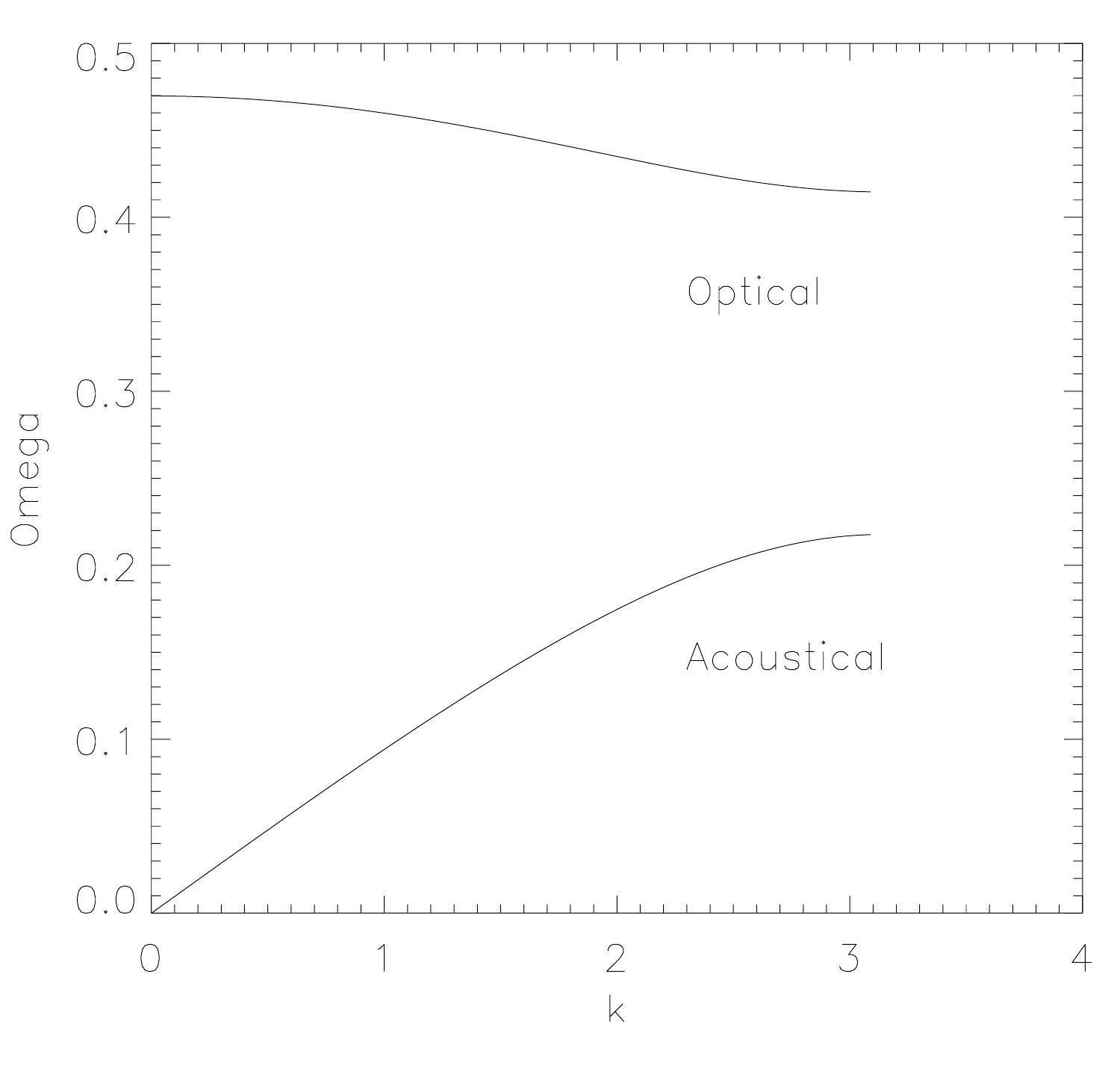}
\caption{\small Graph of $\omega(k)$ against $k$, with $k$ lying in the
  interval [0,$\pi$]. We have chosen $M_{1}$=1,
  $M_{2}$=1.5, $k_{1}$=1, $k_{2}$=0.3 and $\tau$=$10^{-1/2}$.}
\label{fig:frequenza}
\end{center}
\end{figure*}

\begin{figure*}
\begin{center}
\includegraphics[width=8cm]{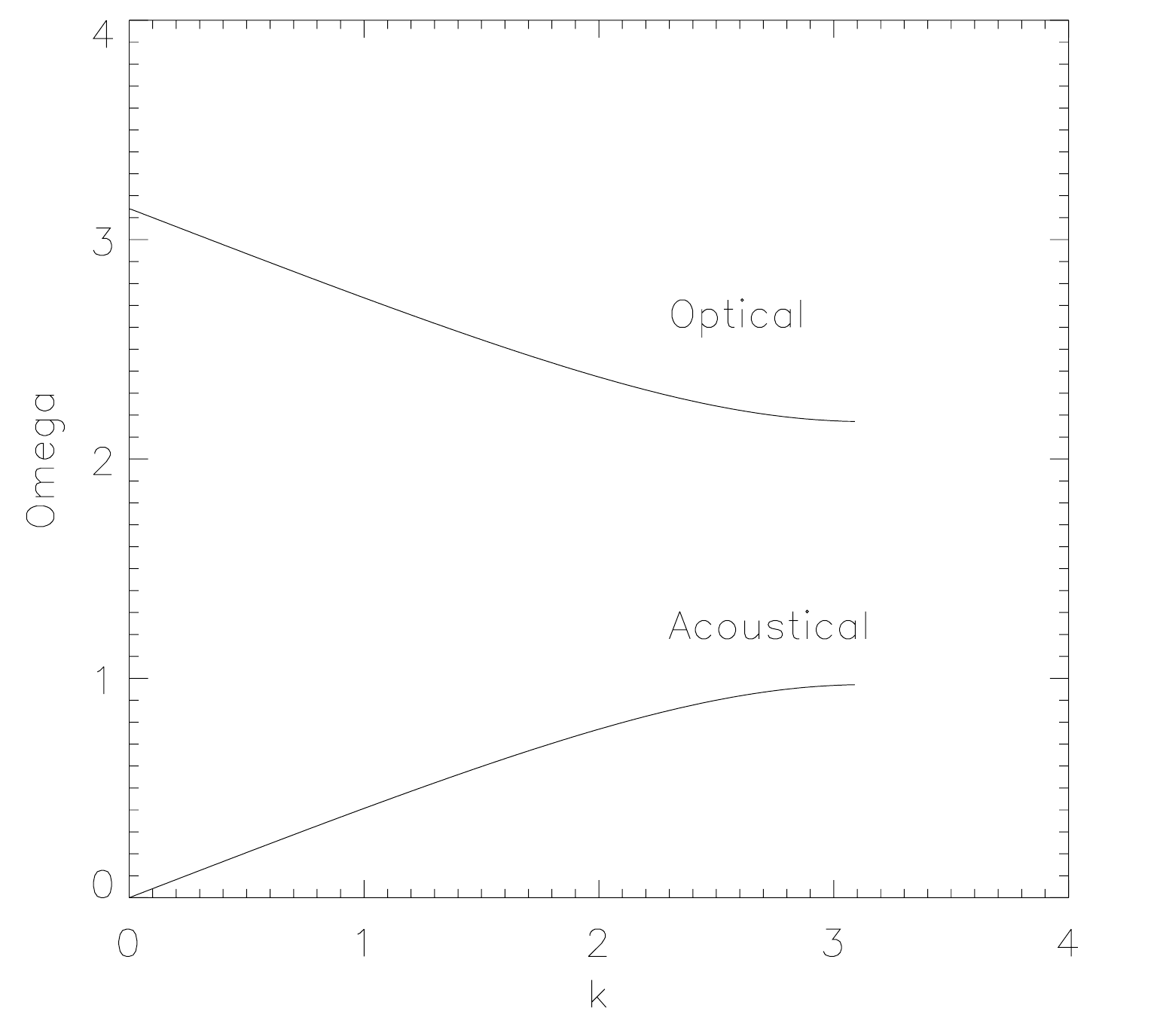}
\caption{\small Graph of $\omega(k)$ against $k$, with $k$ lying in the
  interval [0,$\pi$]. The parameters $M_{1}$, $M_{2}$, $k_{1}$, and $k_{2}$
  are the same as in Fig. 4, but $\tau$=$\tau_{o}$.}
\label{fig:aldo}
\end{center}
\end{figure*}

In equation (\ref{eq:dispersion}), the positive sign corresponds to the
optical branch $\omega_{opt}(k)$, whereas the negative one to the acoustical
branch $\omega_{ac}(k)$.  It can be proved that the function $\omega(k)$ is real
for all real values of $k$ iff the temporal scale $\tau$ satisfies the
following inequalities:
\begin{equation}
  \tau{\leq}\sqrt{\frac{4M_{1}M_{2}}{(k_{1}+k_{2})(M_{1}+M_{2})}}\equiv\tau_{o}
\label{eq:c4}
\end{equation}
for the optical branch, and
\begin{equation}
  \tau{\leq}\sqrt{\frac{8M_{1}M_{2}}{(k_{1}+k_{2})(M_{1}+M_{2})
      -\sqrt{(k_{1}+k_{2})^{2}(M_{1}+M_{2})^{2}
        -16k_{1}k_{2}M_{1}M_{2} }}}   {\equiv}\tau_{a}
\label{eq:c44}
\end{equation}
for the acoustical one. It is easy to show that $\tau_{a}$ is always larger
than $\tau_{o}$.  In Figs. (\ref{fig:frequenza}, \ref{fig:aldo}) we show how
$\omega(k)$ varies as a function of $\tau$. We have chosen, following
Campa\cite{campa} , the following numerical values for the parameters,
$M_{1}$=1, $M_{2}$=1.5, $k_{1}$=1 and $k_{2}$=0.3, so that
{$\tau_{o}$}$\simeq$1.358732 and {$\tau_{a}${$\simeq$}2.910816. So the
  obtained threshold values $\tau_{o}$ and $\tau_{a}$ are consistent with the
  request that $\tau$, the discretization parameter, be smaller than one.
  
  Let us seek a finite amplitude solution of the nonlinear system (\ref{eq:1}, \ref{eq:2}). To do so,
  we write $x_{n,m}$ and $y_{n,m}$ in terms of the harmonics of the linearized
  equation (\ref{eq:onda}) \bea
  x_{n,m}&=&\sum_{s=0}^{\infty}G_{n,m}^{s}(E_{n,m})^{s}+
\label{eq:A}\sum_{s=1}^{\infty}\bar{G}_{n,m}^{s}(E_{n,m})^{-s},
\\
y_{n,m}&=&\sum_{s=0}^{\infty}H_{n,m}^{s}(E_{n,m})^{s}+
\label{eq:B}\sum_{s=1}^{\infty}\bar{H}_{n,m}^{s}(E_{n,m})^{-s},
\eea 
where, as the variables $x_{n,m}$ and $y_{n,m}$ are real, ($\bar
G_{n,m}^{s}$, $\bar H_{n,m}^{s}$)  are the complex conjugates of the modulation
coefficients ($G_{n,m}^{s}$, $H_{n,m}^{s}$).    We  choose
$G_{n,m}^{s}$=$g_{n_{1},m_{1},m_{2}}^{s}$ and
$H_{n,m}^{s}$=$h_{n_{1},m_{1},m_{2}}^{s}$ as slowly varying functions of the
second order in $n_{1}$ and $m_{1}$ and of the first order in $m_{2}$, defined in such a way to avoid secular
terms.
Moreover we expand the functions $g_{n_{1},m_{1},m_{2}}^{s}$ and $h_{n_{1},m_{1},m_{2}}^{s}$ in the small parameter $\epsilon$. So
 we have: \bea
G_{n,m}^{s}&\equiv&\sum_{l=0}^{\infty}\epsilon^{l}g_{n_{1},m_{1},m_{2}}^{(s,l)},
\label{eq:C}
\\
H_{n,m}^{s}&\equiv&\sum_{l=0}^{\infty}\epsilon^{l}
h_{n_{1},m_{1},m_{2}}^{(s,l)}.
\label{eq:D}
\eea

\subsection{Derivation of the equations of motion}

Substituting ansatz (\ref{eq:A}, \ref{eq:B}) into the equations
of motion (\ref{eq:1}, \ref{eq:2}) and taking into account eqs.  (\ref{eq:C}, \ref{eq:D}) we get
 two equations of the form
\begin{equation}
\sum_{s=0}^{\infty}\sum_{l=0}^{\infty}\epsilon^{l}F_{n_{1},m_{1},m_{2}}^{(s,l)}
(E_{n,m})^{s}+\sum_{s=0}^{\infty}\sum_{l=0}^{\infty}\epsilon^{l}
\bar{F}_{n_{1},m_{1},m_{2}}^{(s,l)}(E_{n,m})^{-s}=0,
\end{equation}
where the 
$F_{n_{1},m_{1},m_{2}}^{(s,l)}$ are function only of the slow variables. As
$(E_{n,m})^{s}$ and $(E_{n,m})^{-s}$ are independent functions, its
coefficients must be equal to zero. So  for each power of  $(E_{n,m})$
and  $\epsilon$ we get  sets of equations $F_{n_{1},m_{1},m_{2}}^{(s,l)}=0$ for the slow
varying modulation coefficients $g_{n_{1},m_{1},m_{2}}^{(s,l)}$ and
$h_{n_{1},m_{1},m_{2}}^{(s,l)}$ together with their complex conjugate.

\subsubsection{$\epsilon^0$}
We look here for  the linearized terms. 
In this case, the coefficient of the zeroth harmonic  satisfies the equation
 \begin{equation}
g_{n_{1},m_{1},m_{2}}^{(0,0)}=h_{n_{1},m_{1},m_{2}}^{(0,0)},
\label{eq:eq1}
\end{equation}
whereas the coefficients of the first harmonics gives a set of two equations that are
identically satisfied when $\omega(k)$ satisfies the dispersion relation (\ref{eq:dispersion}) and 
\begin{equation}
\frac{h_{n_{1},m_{1},m_{2}}^{(1,0)}}{g_{n_{1},m_{1},m_{2}}^{(1,0)}}=r.
\label{eq:BdivisoA}
\end{equation}

It can be proven easily that, for {$q$}$\geq$2,
$g_{n_{1},m_{1},m_{2}}^{(q,0)}$=$h_{n_{1},m_{1},m_{2}}^{(q,0)}$=0.

\subsubsection{$\epsilon^1$}
The coefficients of the zeroth harmonic are
 \bea \label{eq:01bis}
  h^{(0,1)}(x_{1},t_{1},t_{2})&=&g^{(0,1)}(x_{1},t_{1},t_{2})
  +\frac{hL_{1}k_{2}}{k_{1}+k_{2}}
\frac{{\partial}g^{(0,0)}(x_{1},t_{1},t_{2})}{{\partial}x_{1}}\\ \nonumber
  &+&\frac{2}{k_{1}+k_{2}}[\beta_{2}|1-e^{-ik}r|^{2}
  -\beta_{1}|1-r|^{2}]|g^{(1,0)}(x_{1},t_{1},t_{2})|^{2},
\eea
or
\bea \label{eq:01}
h_{n_{1},m_{1},m_{2}}^{(0,1)}&=&g_{n_{1},m_{1},m_{2}}^{(0,1)}
+\frac{L_{1}k_{2}}{2(k_{1}+k_{2})}(g_{n_{1}+1,m_{1},m_{2}}^{(0,0)}
-g_{n_{1}-1,m_{1},m_{2}}^{(0,0)})\\ \nonumber
&+&\frac{2}{k_{1}+k_{2}}[\beta_{2}|1-e^{-ik}r|^{2}
-\beta_{1}|1-r|^{2}]|g_{n_{1},m_{1},m_{2}}^{(1,0)}|^{2},
\eea
depending if we use the expansions in terms of derivatives or differences.

For $s$=1 we find a system of two equations in the two unknowns
$g_{n_{1},m_{1},m_{2}}^{(1,1)}$ and $h_{n_{1},m_{1},m_{2}}^{(1,1)}$. This system is 
compatible only if
\begin{equation}
g_{n_{1},m_{1},m_{2}}^{(1,0)}{\equiv}g_{n_{2},m_{2}}^{(1,0)}, 
\label{eq:identidad}
\end{equation}
where $n_2=n_1-m_1$ and \bea \label{eq:condizione2}
h_{n_{1},m_{1},m_{2}}^{(1,1)}=rg_{n_{1},m_{1},m_{2}}^{(1,1)}
+\frac{2i\sin{\omega}m_{1}\omega_{,k} +k_{2}re^{-ik}}{2(k_{1}+k_{2}e^{-ik})}
L_{1}(g_{n_{2}+1,m_{2}}^{(1,0)}-g_{n_{2}-1,m_{2}}^{(1,0)}), \eea where
$\omega_{,k}${$\equiv$}$\frac{\text{d}\omega}{\text{d}k}$
=$\frac{L_{2}}{L_{1}}$, with $L_{1}$ and $L_{2}$ given in Appendix A.1 by eqs.
(\ref{eq:L1}) and (\ref{eq:L2}).  The differential version of eq.
(\ref{eq:identidad}) is
\[g^{(1,0)}(x_{1},t_{1},t_{2}){\equiv}g^{(1,0)}(x_{2},t_{2}),\]
where $x_{2}${$\equiv$}{$h$}$n_{2}$=$h$($n_{1}-m_{1}$)=$x_{1}-{\frac{h}{\tau}}t_{1}$, and 
\bea
h^{(1,1)}(x_{1},t_{1},t_{2})=rg^{(1,1)}(x_{1},t_{1},t_{2})
+\frac{2i\sin{\omega}m_{1}\omega_{,k}
+k_{2}re^{-ik}}{k_{1}+k_{2}e^{-ik}}
hL_{1}\frac{{\partial}g(x_2,t_{2})^{(1,0)}}{\partial{x_2}}.
\label{eq:condizione2bis}
\eea

For the second harmonic we get
\bea
 g_{n_{1},m_{1},m_{2}}^{(2,1)}&=& K_{1}g_{n_{1},m_{1},m_{2}}^{(1,0){2}},
\label{eq:211}
\\
h_{n_{1},m_{1},m_{2}}^{(2,1)}&=& K_{2}g_{n_{1},m_{1},m_{2}}^{(1,0){2}},
\label{eq:212}
\eea
where $K_{1}$ and $K_{2}$ are given in Appendix A.1 by eqs. (\ref{eq:K1}, \ref{eq:K2}).
It can be easily proven that, for {$q$}$\geq$3,
$g_{n_{1},m_{1},m_{2}}^{(q,1)}$=$h_{n_{1},m_{1},m_{2}}^{(q,1)}$=0.

\subsubsection{$\epsilon^2$}
 Taking into account eq. (\ref{eq:identidad}),  the zeroth harmonic gives a system of two equations  that is satisfied only if
\begin{equation}
\begin{array}{c}
  L_{1}^2(g_{n_{2}+1,m_{2}}^{(0,0)}+g_{n_{2}-1,m_{2}}^{(0,0)}-2g_{n_{2},m_{2}}^{(0,0)})
  =L_{1}\frac{c^{0}}{2}(|g_{n_{2}+1,m_{2}}^{(1,0)}|^{2}
-|g_{n_{2}-1,m_{2}}^{(1,0)}|^{2})\\
  +L_{1}\frac{c^{1}}{2}\{g_{n_{2},m_{2}}^{(1,0)}(\bar{g}_{n_{2}+1,m_{2}}^{(1,0)}
  -\bar{g}_{n_{2}-1,m_{2}}^{(1,0)})+\bar{g}_{n_{2},m_{2}}^{(1,0)}(g_{n_{2}+1,m_{2}}^{(1,0)}
  -g_{n_{2}-1,m_{2}}^{(1,0)})\},
\end{array}
\label{eq:++-}
\end{equation}
where $c^{0}$ and $c^{1}$ are two real constants given in Appendix \ref{Constants}. Defining
\bea \label{eq:02+}
A_{n_{2},m_{2}}&{\equiv}&L_{1}(g_{n_{2}+1,m_{2}}^{(0,0)}-g_{n_{2},m_{2}}^{(0,0)})
-\frac{c^{0}}{2}(|g_{n_{2}+1,m_{2}}^{(1,0)}|^{2}
+|g_{n_{2},m_{2}}^{(1,0)}|^2)\\ \nonumber 
&-&\frac{c^{1}}{2}(g_{n_{2},m_{2}}^{(1,0)}\bar{g}_{n_{2}+1,m_{2}}^{(1,0)}+
\bar{g}_{n_{2},m_{2}}^{(1,0)}g_{n_{2}+1,m_{2}}^{(1,0)}),
\eea
eq. (\ref{eq:++-}) reads:
\begin{equation}
A_{n_{2}+1,m_{2}}-A_{n_{2},m_{2}}=0.
\end{equation}
 Thus $A_{n_{2},m_{2}}=C(m_{2})$, where $C(m_{2})$ is an arbitrary function of
 $m_{2}$. Using the fact that $g_{n_{2},m_{2}}^{(0,0)}$ is a slowly varying
 function in $n_{2}$ we have
\begin{equation}
\begin{array}{c}
L_{1}(g_{n_{2}+1,m_{2}}^{(0,0)}-g_{n_{2}-1,m_{2}}^{(0,0)})
=c^{0}(|g_{n_{2}+1,m_{2}}^{(1,0)}|^{2}+|g_{n_{2},m_{2}}^{(1,0)}|^2)\\
+c^{1}(g_{n_{2},m_{2}}^{(1,0)}\bar{g}_{n_{2}+1,m_{2}}^{(1,0)}+
\bar{g}_{n_{2},m_{2}}^{(1,0)}g_{n_{2}+1,m_{2}}^{(1,0)})+C(m_{2}).
\end{array}
\label{eq:02}
\end{equation}
Eq. (\ref{eq:02}) written in terms of the
derivatives reads:
\begin{equation}
\begin{array}{c}
hL_{1}\frac{{\partial}g^{(0,0)}(x_{2},t_{2})}{\partial{x_2}}
=(c^{0}+c^{1})|g_{n_{2},m_{2}}^{(1,0)}|^2
+\frac{C(m_{2})}{2}.
\end{array}
\label{eq:02bis}
\end{equation}
If we transform the derivatives of eq. (\ref{eq:02bis}) into
differences (using again eq. (\ref{e2.5h}), and recalling that
$x_{2}$={$h$}$n_{2}$), we have
\begin{equation}
\begin{array}{c}
L_{1}(g_{n_{2}+1,m_{2}}^{(0,0)}-g_{n_{2}-1,m_{2}}^{(0,0)})
=2(c^{0}+c^{1})|g_{n_{2},m_{2}}^{(1,0)}|^2+C(m_{2}),
\end{array}
\label{eq:02biss}
\end{equation}
 an equation simpler than eq. (\ref{eq:02}). This difference is due to the fact
that  eq. (\ref{eq:02bis}) is obtained using the Leibniz's rule and an integration, while in the case of eq.
(\ref{eq:02}) the Leibniz's rule is not applicable as we deal with differences.

Finally, for $s$=1, we get a system of two equations in the two unknowns,
$g_{n_{2},m_{2}}^{(1,2)}$ and $h_{n_{2},m_{2}}^{(1,2)}$, which is compatible and not--secular 
only if  
\begin{equation}
\begin{array}{c}
iB_{1}(g_{n_{2},m_{2}+1}^{(1,0)}-g_{n_{2},m_{2}-1}^{(1,0)})
+B_{2}L_{1}^{2}(g_{n_{2}+1,m_{2}}^{(1,0)}
+g_{n_{2}-1,m_{2}}^{(1,0)}-2g_{n_{2},m_{2}}^{(1,0)})\\
+B_{3}|g_{n_{2},m_{2}}^{(1,0)}|^{2}g_{n_{2},m_{2}}^{(1,0)}+
\{B_{4}(|g_{n_{2}+1,m_{2}}^{(1,0)}|^{2}+|g_{n_{2},m_{2}}^{(1,0)}|^2)\\+
B_{5}(g_{n_{2},m_{2}}^{(1,0)}\bar{g}_{n_{2}+1,m_{2}}^{(1,0)}+
\bar{g}_{n_{2},m_{2}}^{(1,0)}g_{n_{2}+1,m_{2}}^{(1,0)})+
 B_{6}C(m_{2})\}g_{n_{2},m_{2}}^{(1,0)}=0.
\end{array}
\label{eq:NLS0}
\end{equation}
Here the coefficients $B_{i}$ ($i$=1,...,6) are real and given in  Appendix
\ref{Constants}. This is a  NLSE on
the lattice. At difference from the standard discrete--time NLS equation presented by Ablowitz and Ladik \cite{al}, this is completely local but not integrable \cite{ramani,viallet}. In the development of $x_{n,m}$
and $y_{n,m}$, $g^{(1,0)}_{n_2,m_2}$ is the main term  which multiplies
$\epsilon^{0}$ and $E_{n,m}$.
If we require that $g_{n_{2},m_{2}}^{(s,l)}$ and
$h_{n_{2},m_{2}}^{(s,l)}$ are localized with respect to $n_{2}$,  we have to set $C(m_{2})=0$
and eq. (\ref{eq:NLS0}) becomes
\begin{equation}  \label{eq:NLS}
\begin{array}{c}
iB_{1}(g_{n_{2},m_{2}+1}^{(1,0)}-g_{n_{2},m_{2}-1}^{(1,0)})
+B_{2}L_{1}^{2}(g_{n_{2}+1,m_{2}}^{(1,0)}
+g_{n_{2}-1,m_{2}}^{(1,0)}-2g_{n_{2},m_{2}}^{(1,0)})\\
+B_{3}|g_{n_{2},m_{2}}^{(1,0)}|^{2}g_{n_{2},m_{2}}^{(1,0)}+
\{B_{4}(|g_{n_{2}+1,m_{2}}^{(1,0)}|^{2}+|g_{n_{2},m_{2}}^{(1,0)}|^2)\\+
B_{5}(g_{n_{2},m_{2}}^{(1,0)}\bar{g}_{n_{2}+1,m_{2}}^{(1,0)}+
\bar{g}_{n_{2},m_{2}}^{(1,0)}g_{n_{2}+1,m_{2}}^{(1,0)})\}g_{n_{2},m_{2}}^{(1,0)}=0.
\end{array}
\end{equation}

\section{CONTINUUM LIMIT OF THE DISCRETE NLS}
Eq. (\ref{eq:NLS}) is obtained from eqs. (\ref{eq:1}, \ref{eq:2}) by discretizing the continuous time variable. This discretization was  necessary  to be able to solve the $l=1$, $s=1$ system which otherwise would have been an unsolvable  linear differential difference wave equation. By discretizing we get a discrete wave equation whose general solution is given by an arbitrary function of a discrete  variable. 

It is  interesting to  perform the limit when the discrete time $m_1$ is transformed into a continuous $t$--variable. 
To do so, we take the limit when $\tau$  goes to zero and $m$ tends to $\infty$ in
such a way that the product $\tau m=t$ is finite.  So eq.  (\ref{eq:NLS})  becomes the integrable NLSE
\begin{equation}
iA_{1}\frac{{\partial}g^{(1,0)}(z_{2},t_{2})}{{\partial}t_{2}}
+A_{2}\frac{\partial^{2}g^{(1,0)}(z_{2},t_{2})}{\partial{z_{2}}^2}
+[A_{3}|g^{(1,0)}(z_{2},t_{2})|^{2}+A_{4}C(t_{2})]g^{(1,0)}(z_{2},t_{2})=0,
\label{eq:berluscon}
\end{equation}
where
$t_{2}$=$\lim_{\tau\rightarrow{0}}\lim_{m\rightarrow{\infty}}\tau{m_{2}}$ and
$z_{2}$=$\frac{1}{N}(n_1-\frac{\text{d}\Omega}{\text{d}k} t_1)$ is a new continuous variable. The
coefficients $A_{i}$ ($i$=1,...,4)  in this limit are finite and real, and are given by
\bea \nonumber 
A_{1}&=&\lim_{\tau\rightarrow{0}}2{\tau}B_{1}=-\Omega
\frac{(M_{1}+M_{2})(k_{1}+k_{2})-2M_{1}M_{2}\Omega^{2}}
{k_{1}+k_{2}-M_{2}\Omega^{2}}, \\ \nonumber
A_{2}&=&\lim_{\tau\rightarrow{0}}B_{2}={\scriptstyle \frac{[(M_{1}+M_{2})(k_{1}+k_{2})
-M_{1}M_{2}\Omega^{2})](\Omega_{,k})^{2}
-M_{1}M_{2}(\Omega_{,k})^{2}-k_{1}k_{2}\cos{k}}
{k_{1}+k_{2}-M_{2}\Omega^{2}}},\\ \nonumber
  A_{3}&=&\lim_{\tau\rightarrow{0}}(B_{3}+2B_{4}+2B_{5})
=\lim_{\tau\rightarrow{0}}(B_{3}+2(c_{0}+c_{1})B_{6})\\ \nonumber
  &=&-{\scriptstyle 2\beta_{1}^{2}(\bar{R}-1)\{(R-1)|R-1|^{2}
  \frac{2k_{2}(1-\cos{k})-(M_{1}+M_{2})\Omega^{2}}{D}
  +\frac{2(R-1)}{k_{1}+k_{2}}|1-R|^{2}\}}\\ \nonumber
  &+&{\scriptstyle 2\beta_{2}^{2}(1-\bar{R}e^{ik})\{-(1-Re^{-ik})|1-Re^{-ik}|^{2}
  \frac{2k_{1}(1-\cos{k})-(M_{1}+M_{2})\Omega^{2}}{D}
  +\frac{2(Re^{-ik}-1)}{k_{1}+k_{2}}|1-Re^{-ik}|^{2}\}}\\ \nonumber
  &+&2\beta_{1}\beta_{2}(\bar{R}-1)\{(\bar{R}-1)(1-Re^{-ik})^{2}
\frac{(M_{2}+M_{1}e^{2ik})\Omega^{2}}{D}\\ \nonumber
&+&{\scriptstyle \frac{2(R-1)}{k_{1}+k_{2}}|1-Re^{-ik}|^{2}\}
+2\beta_{1}\beta_{2}(1-\bar{R}e^{ik})\{(R-1)^{2}(1-\bar{R}e^{ik})
\frac{(M_{2}+M_{1}e^{-2ik})\Omega^{2}}{D}}\\ \nonumber
&+&\frac{2(1-Re^{-ik})}{k_{1}+k_{2}}|1-R|^{2}\}+2gA_{4}, \\ \nonumber
A_{4}&=&\lim_{\tau\rightarrow{0}}B_{6}=\frac{k_{1}\beta_{2}|1-Re^{-ik}|^{2}
+k_{2}\beta_{1}|1-R|^2}{k_{1}+k_{2}},
\eea
where
\[g=\lim_{\tau\rightarrow{0}}(c_{1}+c_{2})\frac{2\beta_{2}k_{1}|1-Re^{-ik}|^{2}
+2\beta_{1}k_{2}|1-R|^{2}}
{(M_{1}+M_{2})(k_{1}+k_{2})(\Omega_{,k})^{2}-k_{1}k_{2}},\]
 \begin{eqnarray}
D=[k_{1}+k_{2}-M_{1}\Omega^{2}][k_{1}+k_{2}-M_{2}\Omega^{2}]
-(k_{1}^{2}+k_{2}^{2}+2k_{1}k_{2}\cos{2k}),
\end{eqnarray}
and 
\[R=\lim_{\tau\rightarrow{0}}r=\frac{k_{1}+k_{2}-M_{1}\Omega^2}{k_{1}+k_{2}e^{-ik}}.\]

$\Omega(k)$=$\lim_{\tau\rightarrow{0}}\frac{\omega(k)}{\tau}$ gives back the continuous dispersion relation \cite{campa}.

\section{CONCLUSIONS}  \label{s4}

In this work,  introducing the concepts necessary for applying the
 perturbative multiscale method to discrete equations we have obtained  a rescaled discrete equation.  We have applied this technique to a biatomic chain model. In this way we have shown that we
can perform in a coherent way a multiscale expansion on the lattice. If we
want to remain on the lattice and want to avoid nonlocality then we need to
restrict ourselves to slow--varying functions. This restriction on the class of
function implies that some of the properties of the starting system will be
lost. Among them by sure that of the integrability, which is strictly
related to the analytic properties of the solutions.

 We have found that $g^{(1,0)}$ (the slowly varying coefficient of the first
harmonic) satisfies a totally discrete local version of the discrete NLSE. One interesting feature of our
discrete NLSE is that, when we perform the continuous limit in the time variable,
 the spatial variable becomes continuous, and we get the continuous integrable
NLSE  (\ref{eq:berluscon}) as in the work by
Campa et al. \cite{campa}. \\

\appendix
\section{Explicit Formulas}

\subsection{$g_{n_{2},m_{2}}^{(1,1)}$ and $h_{n_{2},m_{2}}^{(1,1)}$}

Let  us consider the expansion of the equations of motion  with $l$=$s$=1. In this case
we get a system of two equations in two unknowns,
$g_{n_{1},m_{1},m_{2}}^{(1,1)}$ and $h_{n_{1},m_{1},m_{2}}^{(1,1)}$, that is
compatible only if
\begin{equation}
\begin{array}{c}
[(k_{1}+k_{2})(m_{1}+m_{2})+4m_{1}m_{2}(\cos{\omega}-1)]
 \sin(\omega)L_{2}(g_{n_{1},m_{1}+1,m_{2}}^{(1,0)}-\\
-g_{n_{1},m_{1}-1,m_{2}}^{(1,0)})
 +k_{1}k_{2}\sin{k}L_{1}
(g_{n_{1}+1,m_{1},m_{2}}^{(1,0)}-g_{n_{1}-1,m_{1},m_{2}}^{(1,0)})=0.
\label{eq:BB2}
\end{array}
\end{equation}
It is convenient to choose
\begin{equation}
L_{1}=S\sin(\omega)[(k_{1}+k_{2})(m_{1}+m_{2})+4m_{1}m_{2}(\cos{\omega}-1)]
\label{eq:L1}
\end{equation}
and
\begin{equation}
L_{2}=Sk_{1}k_{2}\sin{k},
\label{eq:L2}
\end{equation}
where $S$ is a real number such that $L_{1}$ ($L_{2}$) is an integer number. In terms of $L_1$ and $L_2$ the dispersion relation becomes $  {\omega}_{,k}=\frac{L_{2}}{L_{1}}.$
With this choice of $L_{1}$ and $L_{2}$, and assuming that
$g_{n_{1},m_{1},m_{2}}^{(1,0)}=g_{n_{2},m_{2}}^{(1,0)}$,
with $n_{2}${$\equiv$}$n_{1}-m_{1}$, we find that eq. (\ref{eq:BB2}) is
satisfied. Thus the system of equations we are studying is compatible, and
leads us to the eq. (\ref{eq:condizione2}).

\subsection{The discrete NLSE}

In this Appendix, we show the steps necessary to find the discrete NLSE
(\ref{eq:NLS}). First, we take the equations of motion, and select the
harmonic $s$=1 with $l$=2. In this way we get a system of two equations in the
two unknowns $g_{n_{2},m_{2}}^{(1,2)}$ and $h_{n_{2},m_{2}}^{(1,2)}$, which is
compatible only if the nonhomogeneous first order difference equation
\begin{equation}
\begin{array}{c}
 [(k_{1}+k_{2})(m_{1}+m_{2})+4m_{1}m_{2}(\cos{\omega}-1)]
\sin(\omega)L_{2}(g_{n_{1},m_{1}+1,m_{2}}^{(1,1)}\\
-g_{n_{1},m_{1}-1,m_{2}}^{(1,1)})+k_{1}k_{2}\sin{k}L_{1}
(g_{n_{1}+1,m_{1},m_{2}}^{(1,1)}-g_{n_{1}-1,m_{1},m_{2}}^{(1,1)})\\
=F(g_{n_{2}+1,m_{2}}^{(0,0)},g_{n_{2}-1,m_{2}}^{(0,0)}
,g_{n_{2},m_{2}}^{(1,0)}),
\end{array}
\label{eq:risonanza}
\end{equation} 
is satisfied. Here $F${$\equiv$}$F(g_{n_{2}\pm{1},m_{2}}^{(0,0)},g_{n_{2},m_{2}}^{(1,0)})$
is a given nonhomogeneous term. As the l.h.s. of this equation is the
same as that of eq. (\ref{eq:BB2}) (but with $g_{n_{2}\pm{1},m_{2}}^{(1,1)}$
replaced by  $g_{n_{2}\pm{1},m_{2}}^{(1,0)}$), the terms depending on $g^{(1,0)}$ contained in  $F$ lead
to secular terms for the unknown $g^{(1,1)}$. To avoid secular terms, we must set $F$=0 and eq.
(\ref{eq:risonanza}) gives
$g_{n_{1},m_{1},m_{2}}^{(1,1)}$=$g_{n_{2},m_{2}}^{(1,1)}$.

If we substitute $g^{(0,0)}$ given by eq. (\ref{eq:02}) into 
$F$=0, then this condition will   give  eq.
(\ref{eq:NLS}) written in terms of $g^{(1,0)}$.

\subsection{Constants}\label{Constants}

We give here the expressions of the coefficients appearing in eqs.
(\ref{eq:211}, \ref{eq:212}, \ref{eq:02}, \ref{eq:NLS}):
\begin{enumerate}
\item Eqs. (\ref{eq:211}, \ref{eq:212}).
\bea \label{eq:K1}
 K_{1}&\equiv& \{\beta_{1}(r-1)^2[k_{1}+k_{2}e^{ik}-r(k_{1}+k_{2}e^{-2ik})] \\ \nonumber
&-&\beta_{2}(1-re^{-ik})^2[k_{1}+k_{2}e^{ik}-r(k_{1}e^{2ik}+k_{2})]\}/\{rD\},
 \\ \label{eq:K2}
K_{2}&\equiv& \{\beta_{1}(r-1)^2[k_{1}+k_{2}e^{2ik}-r(k_{1}+k_{2}e^{-ik})]\\
&-&\beta_{2}(1-re^{-ik})^2[k_{1}+k_{2}e^{2ik}
-r(k_{1}e^{2ik}+k_{2}e^{ik})]\}/\{D\},
\nonumber
\eea
where
 \bea
D&=&[2m_{1}(\cos{2\omega}-1)+k_{1}+k_{2}][2m_{2}(\cos{2\omega}-1)
+k_{1}+k_{2}]\\&-&(k_{1}^{2}+k_{2}^{2}+2k_{1}k_{2}\cos{2k}).\nonumber
\eea
\item Eq. (\ref{eq:02}):
\[c^{0}\equiv\frac{-2k_{2}[\beta_{2}|1-re^{-ik}|^{2}+\beta_{1}|1-r|^{2}]}
{(m_{1}+m_{2})(k_{1}+k_{2})(\omega_{,k})^{2}-k_{1}k_{2}},\]
\[c^{1}\equiv\frac{2\beta_{2}(k_{1}
+k_{2})|1-re^{-ik}|^2}{(m_{1}+m_{2})(k_{1}+k_{2})
(\omega_{,k})^{2}-k_{1}k_{2}}.\]
\item Eq. (\ref{eq:NLS}):
\bea \nonumber
B_{1}&=&-\sin{(\omega)}
\frac{(m_{1}+m_{2})(k_{1}+k_{2})+4m_{1}m_{2}(\cos{\omega}-1)}
{2m_{2}(\cos{\omega}-1)+k_{1}+k_{2}},\\ \nonumber
B_{2}&=&{\scriptstyle \frac{[(m_{1}+m_{2})(k_{1}+k_{2})
+4m_{1}m_{2}(\cos{\omega}-1)]\cos(\omega)(\omega_{,k})^{2}
-m_{1}m_{2}{sin}^{2}(\omega)(\omega_{,k})^{2}-k_{1}k_{2}\cos{k}}
{k_{1}+k_{2}+2m_{2}(\cos{\omega}-1)}},
\eea
\bea \nonumber
B_{3}&=&-{\scriptstyle 2\beta_{1}^{2}(\bar{r}-1)\{(r-1)|r-1|^{2}
\frac{2k_{2}(1-\cos{k})+2(m_{1}+m_{2})(\cos{\omega}-1)}{D}}\\ \nonumber
&+&\frac{2(r-1)}{k_{1}+k_{2}}|1-r|^{2}\}\\ \nonumber
&+&{\scriptstyle 2\beta_{2}^{2}(1-\bar{r}e^{ik})\{-(1-re^{-ik})|1-re^{-ik}|^{2}
\frac{2k_{1}(1-\cos{k})+2(m_{1}+m_{2})(\cos{\omega}-1)}{D}}\\ \nonumber
&+&\frac{2(re^{-ik}-1)}{k_{1}+k_{2}}|1-re^{-ik}|^{2}\}\\ \nonumber
&+&{\scriptstyle 2\beta_{1}\beta_{2}(\bar{r}-1)\{(\bar{r}-1)(1-re^{-ik})^{2}
\frac{-2e^{2ik}m_{1}(\cos{\omega}-1)-2m_{2}(\cos{\omega}-1)}{D}}\\ \nonumber
&+&\frac{2(r-1)}{k_{1}+k_{2}}|1-re^{-ik}|^{2}\}\\ \nonumber
&+&{\scriptstyle 2\beta_{1}\beta_{2}(1-\bar{r}e^{ik})\{(r-1)^{2}(1-\bar{r}e^{ik})
\frac{-2e^{-2ik}m_{1}(\cos{\omega}-1)-2m_{2}(\cos{\omega}-1)}{D}}\\ \nonumber
&+&\frac{2(1-re^{-ik})}{k_{1}+k_{2}}|1-r|^{2}\},
\eea 
\bea \nonumber
B_{4}&=&c^{0}B_{6},\\ \nonumber
B_{5}&=&c^{1}B_{6},\\ \nonumber
B_{6}&=&\frac{k_{1}\beta_{2}|1-re^{-ik}|^{2}
+k_{2}\beta_{1}|1-r|^2}{k_{1}+k_{2}}.
\eea
\end{enumerate}

\end{document}